\documentclass[5p,times,twocolumn]{elsarticle}
\pdfoutput=1
\usepackage{graphicx}
\usepackage{xspace}
\usepackage{url}
\usepackage{amssymb}
\usepackage{lineno}

\usepackage[finalnew]{hide/trackchanges} 
\addeditor{sc}
\addeditor{li}
\addeditor{op}
\addeditor{ho}
\addeditor{ak}
\addeditor{cx}

\newcommand{\bhac}{\texttt{BHAC}\xspace}                  \tcregister{\bhac}{1}
\newcommand{\TM}{\texttrademark\xspace}                   \tcregister{\TM}{1}
\newcommand{\R}{\textsuperscript{\textregistered}\xspace} \tcregister{\R}{1}

\newcommand{\footurl}[1]{\footnote{\texttt{\url{#1}}}}    \tcregister{\footurl}{1}
\newcommand{\harm}{\texttt{HARM}\xspace}                  \tcregister{\harm}{1}
\newcommand{\koral}{\texttt{KORAL}\xspace}                  \tcregister{\koral}{1}
\newcommand{\hamr}{\texttt{H-AMR}\xspace}                  \tcregister{\hamr}{1}

\newcommand{\intel}{Intel\xspace}       \tcregister{\intel}{1}
\newcommand{\vtune}{VTune\xspace}       \tcregister{\vtune}{1}
\newcommand{\advi}{Advisor\xspace}      \tcregister{\advi}{1}
\newcommand{\sng}{SuperMUC-NG\xspace}   \tcregister{\sng}{1}

\journal{Astronomy and Computing}

\begin{document}
\begin{frontmatter}

\title{Optimizing the hybrid parallelization of BHAC}

\author[lrz]{Salvatore Cielo}   \ead[URL]{cielo@lrz.de}
\author[uva]{Oliver Porth}      \ead[URL]{o.porth@uva.nl}
\author[lrz]{Luigi Iapichino}   
\author[lenovo]{Anupam Karmakar}
\author[inst1]{Hector Olivares}
\author[inst2]{Chun Xia}

\address[lrz]{Leibniz Supercomputing Centre of the Bavarian Academy of Sciences and Humanities (LRZ), Garching b. 
M{\"u}nchen, Germany}
\address[uva]{Anton Pannekoek Institute for Astronomy, University of Amsterdam, Science Park 904, 1098 XH, Amsterdam, 
The Netherlands}
\address[lenovo]{Lenovo Infrastructure Solutions Group, Lenovo Global Technology, Meitnerstr. 9, D$-$70563 Stuttgart, 
Germany}
\address[inst1]{Department of Astrophysics/IMAPP, Radboud University Nijmegen,
   P.O. Box 9010, 6500 GL Nijmegen, The Netherlands}
\address[inst2]{School of Physics and Astronomy, Yunnan University, Kunming 650050, China}

\begin{abstract}
We present our experience with the modernization on the GR-MHD code \bhac, aimed at improving its novel hybrid 
(MPI+OpenMP) parallelization scheme. In doing so, we showcase the use of performance profiling tools usable on x86 
(Intel-based) architectures. 

Our performance characterization and threading analysis provided guidance in improving the concurrency and thus the 
efficiency of the OpenMP parallel regions. We assess scaling and communication patterns in order to identify and 
alleviate MPI bottlenecks, with both runtime switches and precise code interventions. The performance of optimized 
version of \bhac improved by $\sim28\%$, making it viable for scaling on hundreds of thousands of supercomputer nodes.
 
We finally test whether porting such optimizations to different hardware is likewise beneficial on the new architecture 
by running on ARM A64FX vector nodes.
\end{abstract}


\end{frontmatter}


\section{Overview} \label{s:overview}

Supercomputers have become essential tools in most fields of modern research. Thanks to ever-growing computational resources, the boundary of knowledge and the possibilities of the scientific method have been pushed through otherwise impossible challenges. A very representative example is given by the COVID-19 pandemic: it has been shown that supercomputers are crucial tools for drug discovery and development\footnote{See for instance \url{https://www.compbiomed.eu/} for a description of the efforts in the framework of a current EU-funded consortium in this field.}. Such novel projects have had the additional merit of better introducing supercomputers to the general public, and to funnel new classes of users from industry and science. 
Both novel applications and the ones that are well-established since decades (among the latter, numerical simulations are the main representative ones) have however been facing several challenges in order to utilize at best this powerful resource.

On the verge of the Exascale Computing Era, we can confidently state that scientific applications, and astrophysical 
ones in particular, have --not without significant efforts-- successfully managed several challenges in the field of 
High-Performance Computing (HPC): to keep up with the performance improvement dictated by Moore's law (see 
e.g.~\cite{sutter05freelunch}), new layers of parallelisation have been introduced in the architectures, and programming 
paradigms have been designed accordingly. Examples include the emerging of hybrid (distributed plus shared-memory) 
parallel codes to make efficient use of multicore nodes while allowing good scaling,
even up to full machine size; the introduction of vector registers and the simultaneous development of SIMD (Single Instruction Multiple Data) programming strategies; and structured memory hierarchy of modern processors, to ensure data streaming from the main memory into the computing units. 

Yet the field is undergoing several major breakthroughs, well exemplified by the top places of the TOP500\footurl{https://www.top500.org/} ranking. The usage of GPUs as compute accelerators and the possibilities offered by the latest vector architectures are the most promising technologies in an Exascale perspective. Both of them however present challenges, respectively the necessity of programming the extra devices, and a thorough optimization work. Supercomputing centers are in charge of presenting these technologies to users from the different science fields, foster dialog with vendors and explore viable programming models.

In this work we present our experiences from one such project, the modernization of the \bhac code\footurl{https://bhac.science}, hosted by the Astro- and Plasma- physics Application Lab (AstroLab) of the Leibniz Supecomputing Center in the framework of high-level user support, and operated jointly with the code developers. 

We begin by outlining the code baseline performance and defining the optimization goals (Section \ref{s:goals}). We then characterize the initial code performance (SIMD, MPI layer, roofline analysis) at small and intermediate scales through a first series of scalings and tool-assisted runs (Section \ref{s:profiling}).
In Section \ref{s:results} we instead
describe the main bottlenecks we identify, highlight the implemented code interventions and show the improved performance at scale (Section \ref{ss:final-scaling}).
We finally attempt a performance portability test on a A64FX architecture (Section \ref{s:beast}), before summarizing the optimization roadmap and drawing some general conclusions (Section \ref{s:end}).

\section{Code overview}\label{s:code}

\bhac is an open-source (GPL v3.0) general relativistic MHD code  \cite{PorthOlivares2017,OlivaresPorthEtAl2019}, written in \texttt{Fortran90}, parallelized using \texttt{MPI} and is based on the \texttt{MPI-AMRVAC} Toolkit\footurl{http://amrvac.org} v1.0 \cite{KeppensMeliani2012,PorthXia2014,XiaTeunissenEtAl2018}. 
\bhac offers a variety of numerical methods and fully adaptive block based (bi-, quad- or oct-) tree adaptive mesh refinement (AMR) using staggered mesh constrained transport in any covariant coordinate system \cite{OlivaresPorthEtAl2019}.  
The high-resolution shock-capturing finite volume scheme uses 
{second} order Implicit/Explicit timesteppers combined with 
{second or third} order (PPM) spatial reconstruction.  
\bhac uses robust multi-dimensional non-linear solvers (Newton-Raphson and Newton-Krylov) for the conversion from the integrated conserved variables to the primitive variables with well tested backup strategies \citep[e.g.][]{RipperdaBacchiniEtAl2019}. This allows to perform simulations of challenging 
{magnetically-dominated} regimes as present in pulsar magnetospheres, black hole accretion and relativistic reconnection. 

\bhac has been applied (among other codes like \harm, \koral and \hamr) to build GRMHD models for the simulation library used for the interpretation of the first ``picture of a black hole'' by the Event Horizon Telescope Consortium \cite{CollaborationAkiyamaEtAl2019}. The veracity of GRMHD codes like \bhac has been established in a community spanning code comparison effort \cite{PorthChatterjeeEtAl2019a}.  In \cite{2020MNRAS.497..521O}, \bhac simulations were employed to show how the accretion process onto horizon- and surfaceless ``black hole mimickers'' called Boson stars differs from the black hole case.  The study showed how the EHT can be used to distinguish between different classes of compact objects and thus affirm the black hole hypothesis.  

Numerical simulations with \bhac are also used to unravel the origin of ``flares'' which are observed from the galactic center black hole SgrA* in the X-ray and infrared bands \cite[e.g.][]{Dodds-EdenPorquet2009,GravityCollaborationAbuterEtAl2018}.  Two distinct physical scenarios are currently being investigated \cite{NathanailFrommEtAl2020,2020ApJ...900..100R,2021MNRAS.502.2023P}.  

\bhac is also used to model jet formation in the context of the binary neutron star merger event GW170817 \cite{AbbottAbbott2017}.  In \cite{NathanailGillEtAl2020,Nathanail2021}, \bhac simulations showed, for the first time, how self-consistently launched magnetized outflows provide accurate fits to the multi-frequency afterglow lightcurve of GW170817, allowing to deduce key parameters of the source.  

Yet another research line uses \bhac to study magnetic reconnection in the relativistic regime.  Capitalizing on the AMR capabilities, \textit{effective} resolutions of $8000^3$ have already been achieved which allows to resolve both the large scale plasma instabilities as well as the small scale current sheets forming in the non-linear evolution (e.g. \cite{RipperdaPorth2017a} and Ripperda et al. in prep). 

\section{Initial performance and optimization goals}\label{s:goals}

In the past few years, \bhac has undergone previous modernization projects within LRZ. These mainly targeted a refactoring of the solver scheme towards a task-based algorithm, which also allowed the addressing of bottlenecks in the memory access pattern. Hence the refactoring of important loops of the source code, finally complemented by further efforts towards parallel I/O challenges and SIMD usage. 

\begin{figure}[h]
\includegraphics[width=\columnwidth]{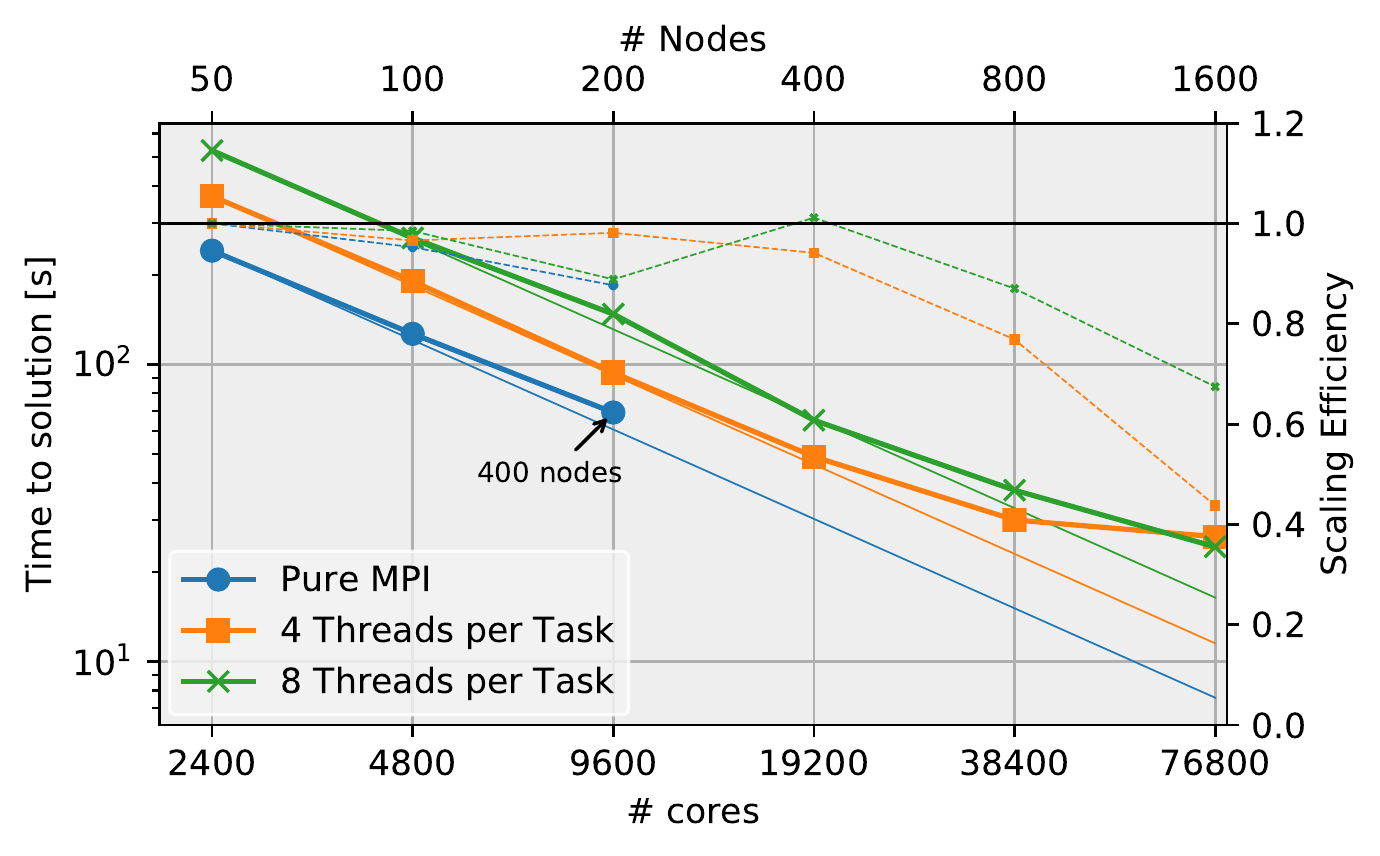}
\caption{Initial scaling on SuperMUC-NG. 
{At 9600 cores, the pure MPI case requires in reality 400 nodes in order not run out of memory, due to the MPI buffer overhead}. Hybrid MPI/OpenMP alleviates this problem, running successfully up to 76800 cores.  However, there is a severe performance penalty when increasing the number of OpenMP threads per task.}
\label{fig:initial-scaling}
\end{figure}

Since then, the code has seen the introduction of a basic hybrid \texttt{MPI}/\texttt{OpenMP} parallelization, which has not yet been used in production, but has proven very promising in extending the scalability range of the code. 
Typical use cases so far have utilized up to 8000 cores, e.g. on SuperMUC, Cartesius/Netherlands or Breniac at VSC/Belgium.  However, we have observed in previous scaling runs that {the MPI overhead of pure MPI runs becomes prohibitive at $\sim 10\, 000$ processes}.  This is demonstrated in Figure \ref{fig:initial-scaling} which shows the \bhac strong-scaling on SuperMUC-NG (48 cores per node). The pure MPI run with 9600 processes in fact allocated 400 nodes, loaded with only half the processes to fit memory requirements.  

Our main target machine is LRZ's SuperMUC-NG, which consists of 6,336  Intel Xeon Skylake Platinum 8174 processor compute nodes each with 48 cores and 96 GB memory (thin nodes), as well as 144 nodes of the same processors each 48 cores and 768 GB memory per node (fat nodes). An Intel Omni-Path (OPA) high-performance fast interconnect provides up to 100G network bandwidth on these nodes. These compute nodes are bundled into 8 domains (islands).The OmniPath network topology within the islands is a 'fat tree' for highly efficient communication and the interconnection between the islands is pruned with a pruning factor 1:4.

The preliminary exploration shown in Figure \ref{fig:initial-scaling} shows the potential of the hybrid parallelization: already with four or eight threads per task, the code is able to run beyond 400 nodes (tested up to 1600 nodes/76800 cores) and scaling is efficient until 800 nodes. However, in its initial state, the hybrid implementation comes with a severe performance penalty, e.g. at eight threads per task, the runtime is twice the pure MPI case at the same node count.

To efficiently run large problems beyond $2000^3$ cells as needed for example in the study of relativistic turbulence, it is necessary to go beyond $10\, 000$ cores accessible with pure MPI parallelization.  Goal of this AstroLab project is hence to understand and improve the performance in the hybrid implementation.  Since this is the first investigation of its kind with the \texttt{MPI-AMRVAC} toolkit, we use a simple uniform grid setup and aim for a fairly complete characterization of the code infrastructure.

{In order to explore all the code bottlenecks, at node level and beyond, we need to run \bhac at different scales, and thus test problems of different sizes.}

We target single node, 16 nodes and 800 nodes, 
\change[sc]{such that we obtain 65536 cells per core in each case, in order to keep the workload per node constant.}
{scaling up the same physical problem while keeping a constant workload of 65536 cells per core.}  Thus our problems range from $\sim 3\times 10^6$ to over $2\times 10^9$ computational cells.  

\section{Profiling of the initial version}\label{s:profiling}

\subsection{single node characterization}
\change[sc]{This task was entirely executed on}
{For the first general profiling runs we used SuperMUC-NG at LRZ.} As the OpenMP implementation was experimental, 
\change[sc]{the correctness check we run} 
{we ran a correctness analysis with the Intel Inspector tool; this check}
yielded several race conditions
\change[sc]{which all could be fixed.}
{which we needed to address before attempting any optimization, but were very easily corrected.} 
At this point we performed a few single node scaling tests and tool-assisted profiling runs to determine the optimal node configuration for the  (mainly compiler-assisted vectorization and MPI/OpenMP ratio). The tools are more properly discussed in Section \ref{s:results}, so for now we just summarize these early results of ours.

Thanks to the previous modernization interventions, \bhac showed good SIMD efficiency, to the point that it was already convenient for us to start with the Xeon's full \texttt{AVX512} instruction set and 512-bit \texttt{ZMM} register, by automatic compiler vectorization (i.e. compiling with the flags {\verb|-xAVX512 -qopt-zmm-usage=high|}).
A $32\%$ speedup from SIMD alone was still theoretically possible; however not very profitable, as the setup showed a prevalence of memory-bound kernels.

The hotspot analysis yielded no pressing need for action, suggesting that we could proceed with the optimization of the OpenMP threading efficiency.
On this respect, node configurations with a high numbers of thread (8 to 32 per node) had actually higher parallel efficiency than with fewer threads (at least for the single node case); thus it was more convenient to run any blind test with such configurations.

This fact also strongly suggests that the problems are likely due the MPI/OpenMP interaction (as using less tasks per node is always more efficient when OpenMP is used). We later (see Section \ref{s:final-scaling}) found this issue to be key, the exact cause residing in the details of MPI's shared memory intra-node communications. A better characterization
of the MPI structure of the code was thus necessary.

\subsection{MPI profiling of the 16 nodes test case} \label{ss:mpi}

Using \emph{mpi$_{-}$trace}
\add[sc]
{(Lenovo's lightweight and extended MPI tracing tool)}, we evaluated the pure MPI scaling of the 16 node test case in the range from 80 to 3200 MPI tasks, in a strong-scaling fashion. These runs were performed at the Lenovo Benchmark Cluster in Stuttgart (``Lenox'') on Lenovo SD530 platform with dual-socket Intel Xeon Gold 6148 based nodes. These nodes have roughly 80 GB available physical memory for the running applications. An Intel Omni-Path (OPA) high-performance fast interconnect provides up to 100G network bandwidth on these nodes. For these baseline measurements Intel fortran compiler version 20.2 and Intel MPI 2019.8 were used. 

\begin{figure}[h!] \begin{center} \includegraphics[width=\columnwidth]{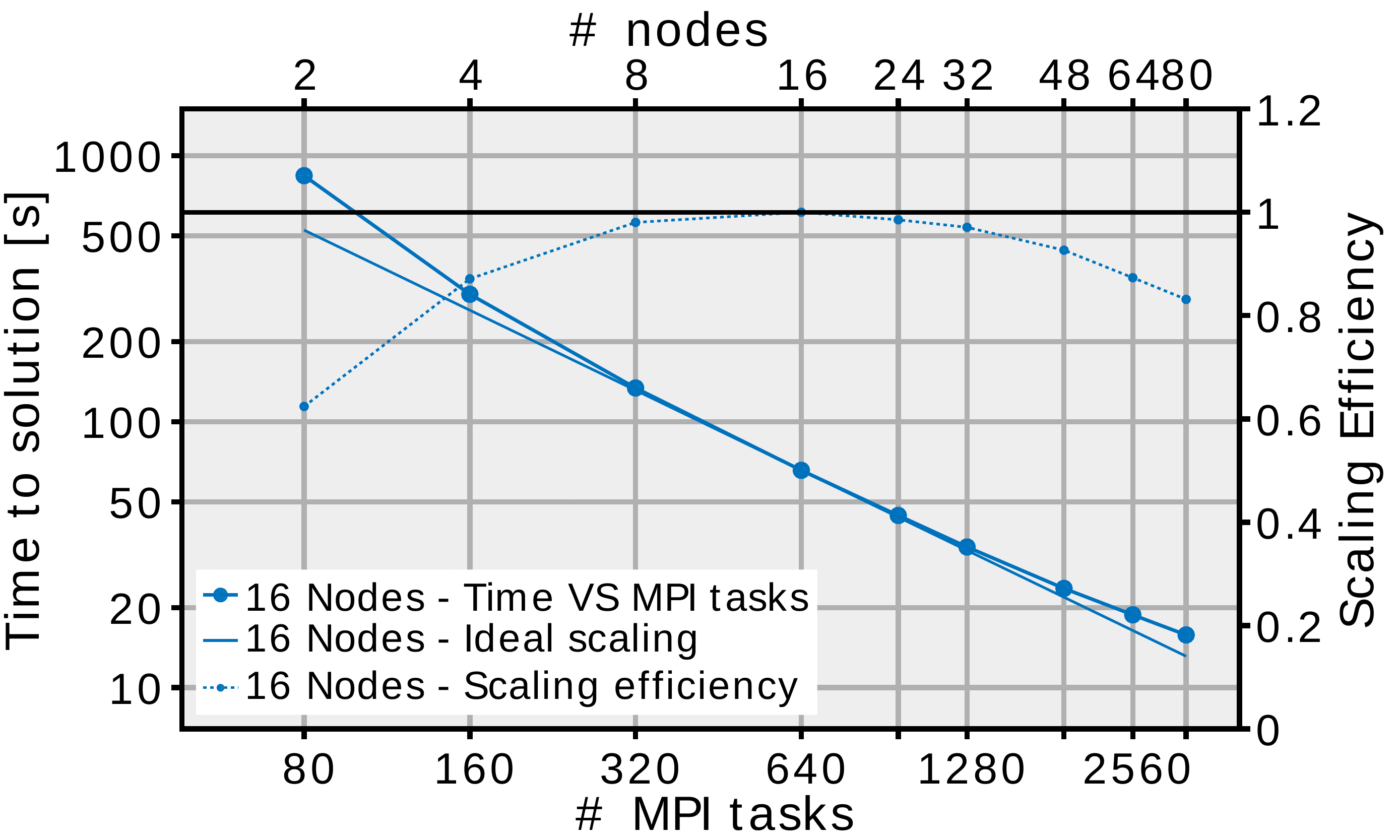}
\caption{Pure MPI scaling (solid lines) of the 16 node use-case on the Lenox cluster. The corresponding scaling efficiency declines in both direction from the designed optimal load of 16 nodes.} 
\label{fig:MPIscaling}
\end{center} \end{figure}

The results obtained are summarized in Figure \ref{fig:MPIscaling}  which shows an excellent strong scaling of up to a factor of $19$ from 160 to $3200$ processes. The run at $80$ MPI tasks ($2$ nodes) significantly under-performed, likely due to caching effects. This experiment shows that the pure MPI implementation using non-blocking \texttt{ISEND/IRECV} is quite efficient, requiring no urgent improvements. 

\begin{figure}[h!] \begin{center} \includegraphics[width=\columnwidth]{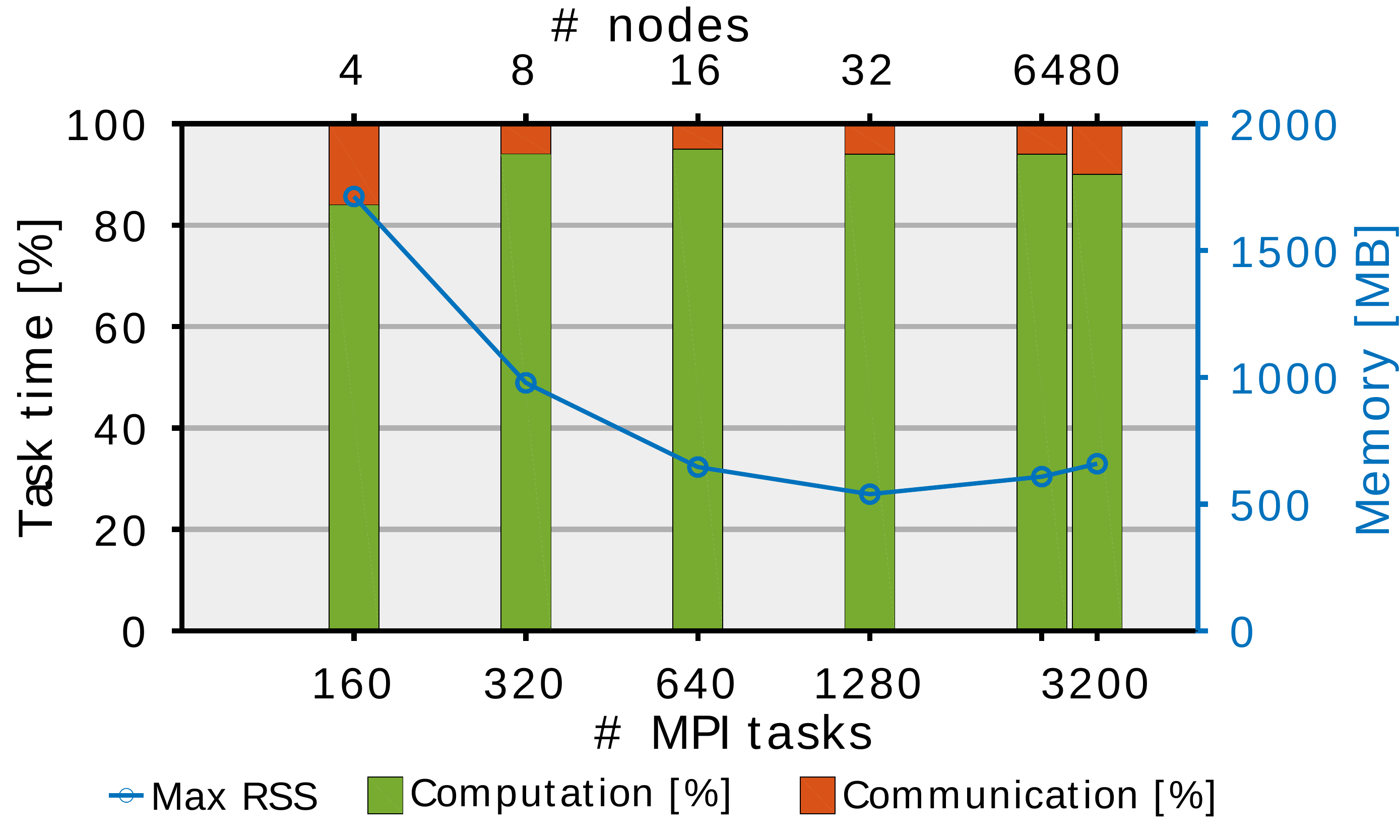}
\caption{Computation/communication fraction (histogram, stacked bars) and memory footprint (solid line) for the 16 nodes, pure MPI setup (from trace analysis on the Lenox cluster). Once more the optimal computation, communication balance and memory requirement per task are achieved for the 16 nodes run.}
\label{fig:MPItrace}
\end{center} \end{figure}

An analysis of the mpi$_{-}$trace findings, Figure \ref{fig:MPItrace}, shows how the relative time spent on MPI communications alters very slowly with the number of MPI tasks, whereas the memory usage rapidly drops due to greater parallelism. This, in a way, explains the good scalability behaviour achieved with node scaling of this use-case.

\begin{figure}[h!] \begin{center} \includegraphics[width=\columnwidth]{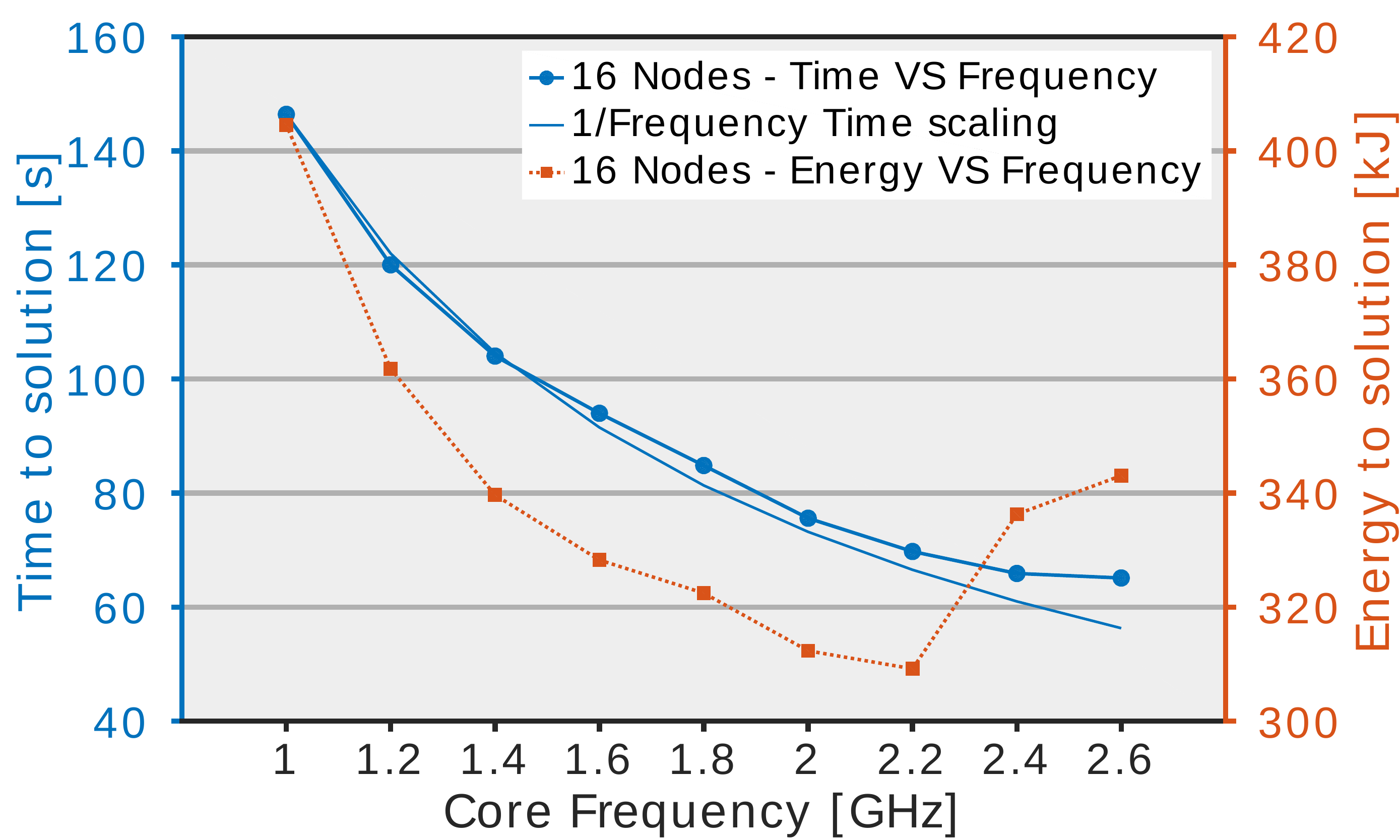}
\caption{Frequency scaling (time to solution, light blue line) of the 16 node pure MPI test setup. The corresponding global energy to solution (orange line) has its clear minimum at a clock frequency of $2.2\rm GHz$.}
\label{fig:MPIfrequency}
\end{center} \end{figure}

As an aside, the same setup was also used to compute the frequency scaling and energy efficiency of the code.  Using 16 nodes, the frequency was scaled from 1 $\rm GHz$ to $2.4\rm GHz$ and the time-to-solution and average node power (Watt) was recorded. The results are illustrated in Figure \ref{fig:MPIfrequency}.

{At an average of $277 \rm W$ per node, the minimum energy-to-solution $(\sim309~\rm kJ)$ was found for a clock speed of $2.2 \rm GHz$, this is an optimum trade-off between performance and energy consumption. At a maximum clock speed, the energy-to-solution increases around $9\%$, however, it slightly}
{lowers} ($\sim 6\%$) the time-to-solution performance. 
{S}witching on the turbo frequency does not positively impact the overall performance, but only surges the energy to solution requirement.

With these numbers in hand, we can compute useful conversion factors indicating the energy and carbon footprint of full numerical simulations with \bhac.  Since the majority of the electricity consumption in astrophysical institutes originates from the use of high-performance computing \citep[e.g.][for the recent analysis of a representative case]{Jahnke2020}, it is important to be sensitive to the power- and climate- impact of numerical simulations.  As argued by \cite{PortegiesZwart2020}, computing related greenhouse gas emissions also surpass telescope operations in astronomy.  We estimate the carbon footprint of simulations performed on hardware with similar characteristics to Lenox as follows:  
A valid metric for the work of grid-based codes such as \bhac is the number of the performed cellupdates, whereby we count each sub-step of the temporal integration multiplied by the total number of cells in the simulation.  
Our test-setup performed $\simeq5.3\times10^9$ individual cellupdates which yields an  energy-per-million-cellupdate conversion of $\rm Epc_6 = 58 \frac{\rm J}{10^6 \rm cellupdates}=1.6\times 10^{-5}\frac{kWh}{10^6 \rm cellupdates}$.  Adopting the average C02-equivalent emission per kWh electrical energy in Europe (2019) \footurl{https://www.eea.europa.eu/ds_resolveuid/7e1d0edbc6b4488a89fb1686d8239858} of $275 \rm \frac{gC02e}{kWh}$, we obtain the emission coefficient $C02_6=4.4\frac{\rm mg}{10^6 \rm cellupdates}$.  
To put this in perspective: for a typical production run within the simulation library in \cite{CollaborationAkiyamaEtAl2019d}, we perform $2\times 10^6 $ iterations, and simulate a grid of $6\times 10^6$ cells.  Using a two-step timeintegrator, the total simulation hence performed $24\times 10^{12}$ cellupdates. Adopting the characteristics found for the test run on Lenox, this corresponds to an energy to solution of $384\rm kWh$ and a C02-equivalent of $106\rm Kg$.  The same amount of greenhouse gasses will be emitted by a 820 km drive in an average new car in Europe (2019) \footurl{https://www.eea.europa.eu/data-and-maps/data/co2-cars-emission-18}.

\subsection{Roofline analysis} \label{ss:roofline}

\begin{figure*}[p] \begin{center} \includegraphics[width=\textwidth]{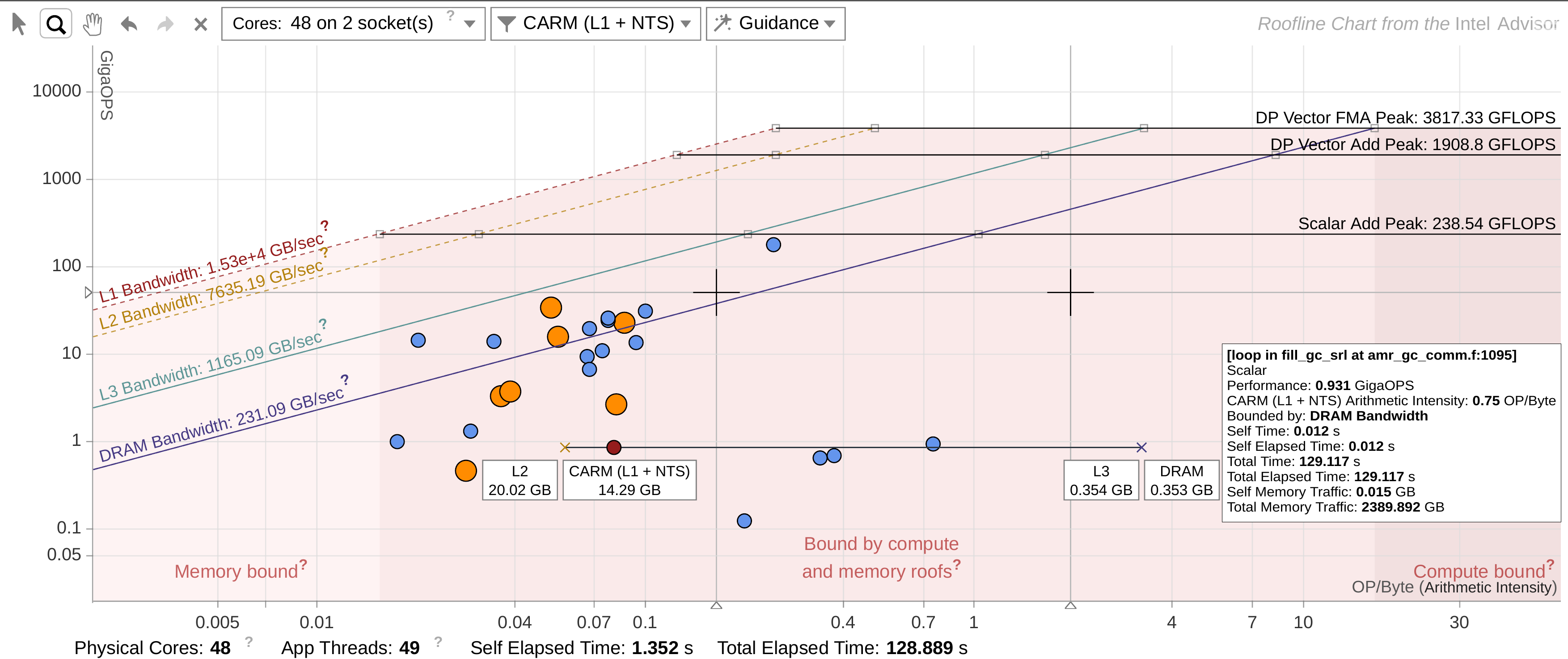}
\caption{VTune roofline analysis. The profiling shows a mixture of compute-bound and memory-bound kernels, with a prevalence of the latter. Such memory issues also hinder the SIMD efficiency of vectorized kernels (in orange), which do not particularly stand out from serial ones (in light blue). Point information for a relevant kernel are shown in the inlet text panel (on-click feature of Advisor).}
\label{fig:roofline}
\end{center} \end{figure*}

Using Intel's Application Performance Snapshot\TM (APS) and Advisor\TM, the performance bottlenecks in the hybrid implementation were identified. Advisor is especially useful in the characterization of the node-level performance (safe for threading utilization, see below), SIMD leverage and memory access efficiency. This information is best summarized in the \emph{roofline analysis}, such as the one we present in Figure \ref{fig:roofline} for the baseline version of \bhac. 

In this plot, the performance of each individual relevant kernel is measured at runtime and plotted against the nominal hardware capabilities, in terms of total FLOPS executed (y-axis) versus \emph{arithmetic intensity} (FLOPS executed per byte, i.e. a measure of how efficiently the memory hierarchy is used; x-axis). Kernel bottlenecks here appear as \emph{roofs}, either horizontal (compute bottlenecks) or slanted (memory bottlenecks). 
Larger circles represent longer runtimes, SIMD kernels are in orange, scalar ones in light blue.

In the figure we also display the information Advisor provides on click (position in the source code, most relevant roof, arithmetic intensity extrema for all memories in the hierarchy, performance figures), for one of the most relevant kernels in the optimization, \verb|fill_gc_srl| (in red, with bold border). 

The analysis shows a mixture of memory-bound and compute-bound kernels, the former being however prevalent. Since memory utilization also limits SIMD efficiency, even that one has room for improvement, confirming the APS results, and in concordance with the fact that vectorized kernels do not appear on average higher than serial ones.

\section{Optimization achievements} \label{s:results}

\subsection{Threading overhaul} \label{ss:vtune}
For the proper OpenMP analysis, the most useful tool provided by Intel is instead VTune Profiler\TM. Its summary view reported an OpenMP imbalance which could be solved by adding \texttt{dynamic} scheduling to the loops with significant workload. This yielded an average performance improvement of $\sim 5\%$.  

Besides this clear summary of the threading utilization, VTune is capable of showing detailed core utilization histograms over time. Idle times can be further distinguished in MPI wait times, OpenMP spin times, or serial OpenMP regions. The graphs can be arranged in several useful ways, and all kernel present in a given time interval are shown with relevant metrics.

\begin{figure*}[p]
\begin{center}
\includegraphics[width=\textwidth]{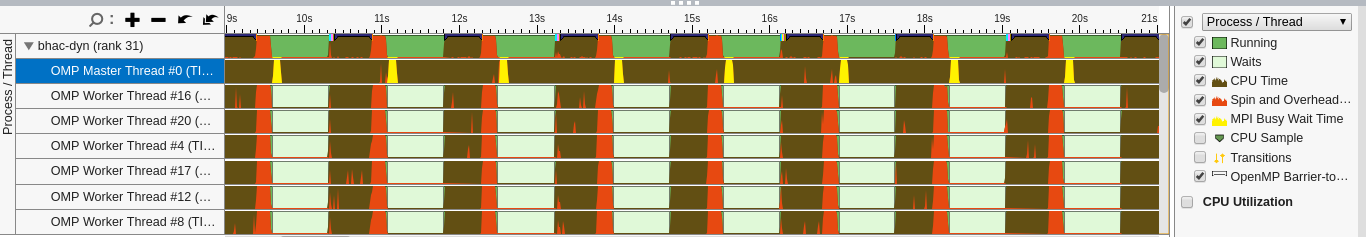}\\ 
\includegraphics[width=\textwidth]{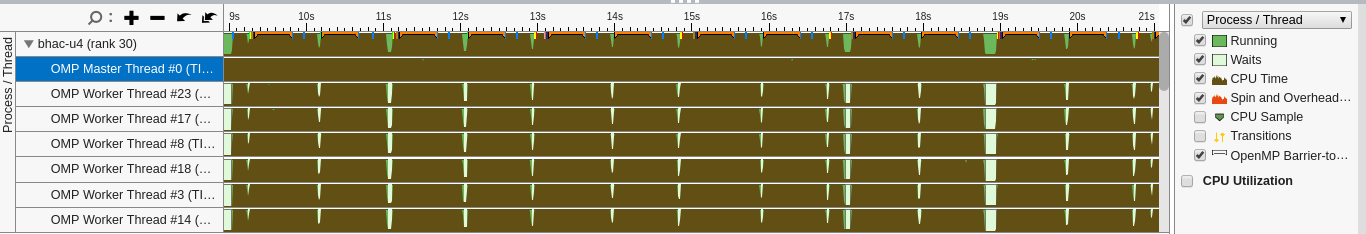}
\caption{Timeline histograms of initial (top) and optimized (bottom) version, from VTune's Bottom-up view. 2 MPI tasks and 24 OpenMP per run. Only a few worker threads and one task/master thread (in blue) are shown. Wait and overhead time (see color legends) are enormously reduced as a consequence of parallelization of the ghost cell exchange region. }
\label{fig:vtune}
\end{center}
\end{figure*}

An example of this is analysis is shown in Figure \ref{fig:vtune}: the considered run featured 2 MPI task per node and 24 OpenMP threads per task; the utilization histograms refer to a single task, with the master thread on top, followed by the other threads that depend on it (for brevity, only 6 working threads are shown). 

Let us at first examine the top panel, which refers to the initial version. The diagrams shows clearly several idle regions, where all worker threads wait for the master to finish long serial calculations and MPI calls (see color legend in the figure). The parallel regions are always well aligned as a consequence of the dynamic scheduling, which improved work-sharing (although with the small impact we mentioned).

{VTune's threading} analysis identified these large serial code blocks within kernels managing the ghost-cell exchange. 
A look at the source and assembly code, also through VTune's preposed interface, 
identified a major OpenMP serial region in the ghost-cell exchange.  The code could then be optimized by the developers via separating out the MPI-dependent and MPI-independent ghostcell operations.  This optimization positively impacts performance in two ways: 1. the MPI-independent parts have been fully OpenMP parallized and 2. the newly OpenMP parallelized operations are executed before the \texttt{MPI-WAITALL} calls of the MPI-dependent operations, allowing more overlap of computation and communication.  

After the refactoring, we repeated VTune's threading analysis (shown in the bottom panel of Figure \ref{fig:vtune}). The idle and wait times are very much reduced, resulting in much more compact parallel regions and thus highly increased core utilization.

\begin{figure}[!]
\begin{center}
\includegraphics[width=\columnwidth]{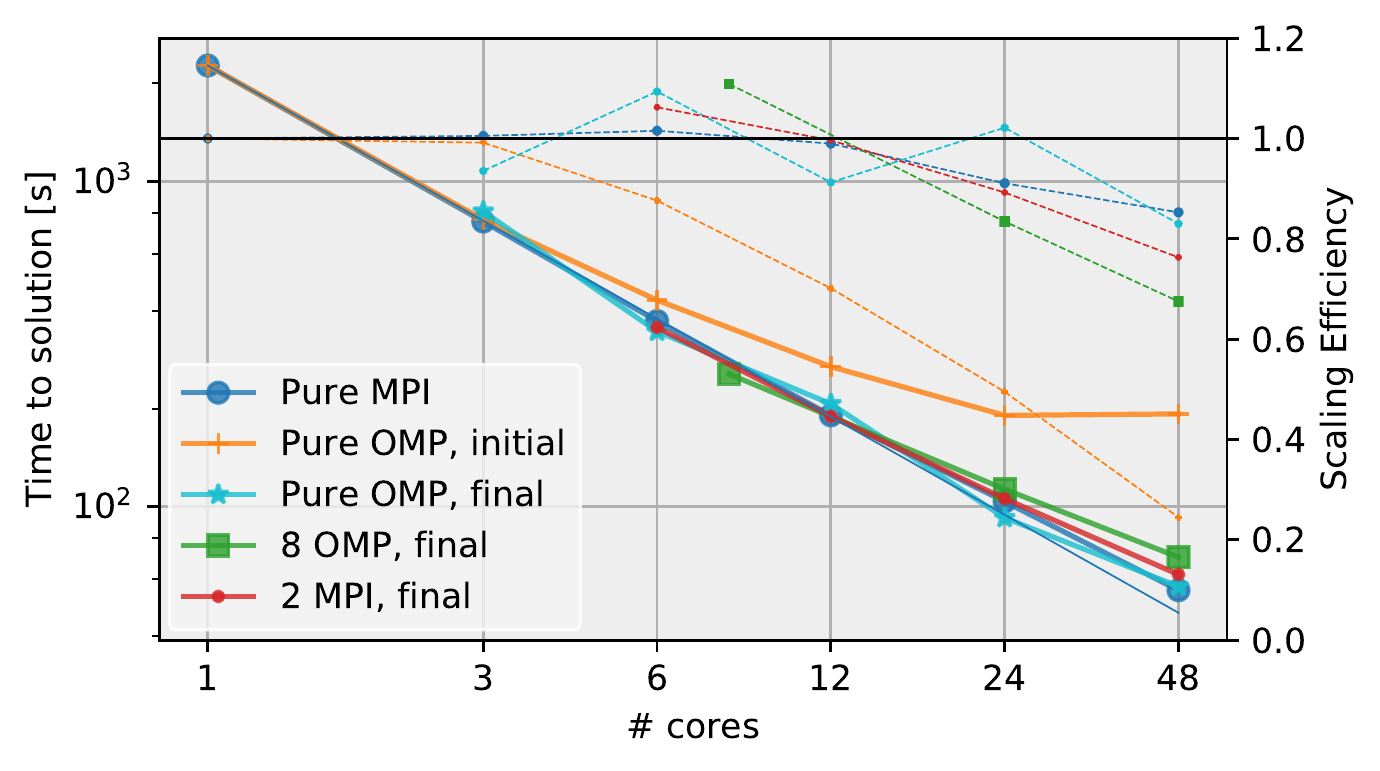}
\caption{Single node scaling with varying contributions of MPI and OpenMP.  Code restructuring has yielded a significant improvement in the pure OpenMP scaling, on par with the pure MPI case.  The curve labeled as ``8 OMP, final'' uses the optimized code with fixed eight threads per task, varying the number of tasks.  The curve ``2 MPI, final'' fixes 2 MPI tasks and investigates scaling by increasing the number of threads.}
\label{fig:single-node}
\end{center}
\end{figure}

Figure \ref{fig:single-node} shows the single node scaling before and after the optimization. While pure MPI scaling (dark blue lines) is close to ideal, the initial pure OpenMP scaling (orange lines) was poor due to the increasing contribution of serial regions (Amdahl's law).  The updated OpenMP scaling (light blue) performs as well as pure MPI.
The hybrid performance was investigated by varying the OpenMP/MPI contributions. In the figure we exemplify this by either choosing 8 OpenMP threadsand increasing MPI tasks, or by fixing two MPI tasks and increasing the OpenMP thread count
\add[sc]
{(see legend)}.  This shows a generally acceptable scaling but also points out room for improvement when mixing MPI and OpenMP.  
\add[sc]
{The optimized OpenMP runs show a clearly improved scaling efficiency (in a few cases also above unity) which testifies the effectiveness of our interventions.}

\subsection{Final performance at scale} \label{ss:final-scaling}

\begin{figure}[!]
\begin{center}
\includegraphics[width=\columnwidth]{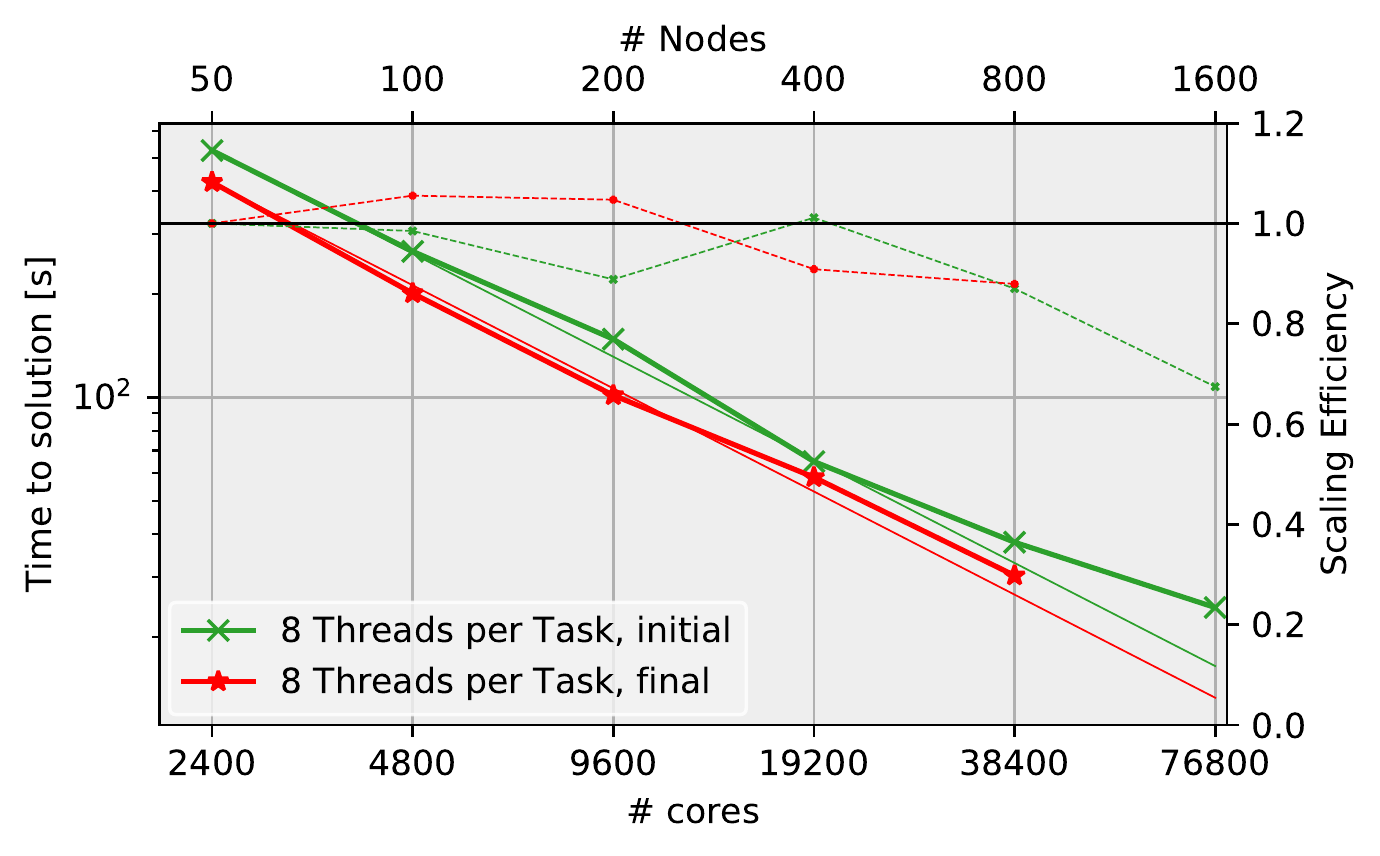}
\caption{Comparison of the initial and final speedup for the large scale setup. The green curve is identical to the one shown in Figure \ref{fig:initial-scaling}.  }
\label{fig:final-scaling}
\end{center}
\end{figure}

Using the updated code version, the initial large scaling run was repeated using 8 threads per task.  The result is shown in Figure \ref{fig:final-scaling}.  
\add[sc]
{Even on this scale, }the data shows that improving the OpenMP parallelism led to an average performance increase of 27\% compared to the initial code.  This shows a significant reduction of the hybrid penalty.  Trace analysis at 16 nodes has indicated that further modifications of the ghost-cell exchange for mixed MPI/OpenMP could improve this even more. Unfortunately
{a know bug of the default MPI version of SuperMUC-NG at the time of writing made pursuing a scaling to 1600 nodes and beyond not worth the effort. Yet the data collected so far give already enough directions for future optimizations.}

{Further exploration of the options offered by Intel MPI was very profitable: switching off \emph{shared-memory intranode communication} by setting}
\verb|export I_MPI_SHM=off| 
{at runtime, we were able to efficiently port the single node speedups to a higher node counts (namely, a speedup of 
$28\%$). Finally we signal that switches to reduce the MPI memory footprint are also available, should the initial 
overhead problem be presented again (e.g. the family of \verb|I_MPI_SHM_CELL_ | and 
\verb|I_MPI_SHM_HEAP_| environment variables, and the control on \verb|I_MPI_MALLOC|), which may also have positive impact on the performance.
}

\section{Test run on A64FX architecture}\label{s:beast}

We finally decided to explore whether our optimization work, so well-performing to Intel x86 architecture, offers some reward when the code is ported to different hardware.

Our target architecture for this experiment were the HPE CS500 nodes of LRZ's BEAST (for \emph{Bavarian Energy, Architecture, and Software Testbed}) cluster \footurl{https://www.lrz.de/presse/ereignisse/2020-11-06_BEAST/} mounting A64FX processors. The A64FX is the 64-bit ARM architecture microprocessor designed by Fujitsu, in force of Fugaku\footurl{https://www.r-ccs.riken.jp/en/fugaku/project} the 537-TFlops supercomputer number one in the Top500 list at the time of writing (since June 2020). The processor is the first to feature the Armv8.2-A instruction set Scalable Vector
Extension (SVE), capable of major performance enhancements on vectorizable operation. The (RISC) SVE -- vectorizing operations of arbitrary length with a single instruction -- operates on different principles than the (CISC/mixed) SIMD extension discussed so far, although the two have the same basic requirements (mainly, avoidance of vector dependencies within loops), and the fact that SIMD versions of some instructions are nevertheless present on the A64FX. It is thus legitimate to wonder how code optimized for one extension performs on the other. Proven benefits of SVE, besides high performance, include improved energy efficiency and ease of manufacturing; all valid motivational reasons which address the core bottlenecks of HPC and scientific computing for the Exascale Era.

During the setup of \bhac a few build issues occurred, for which fixes had to be found. We used Cray \texttt{ftn} compiler v.10.0.1, in an instance containing support for SVE but not for accelerators (not present anyways on the nodes). The \texttt{explain} command was rather useful in clarifying compilation errors, and made the porting process smoother. Nonetheless, compilation macros were adjusted to provide \texttt{.f90} extension to all source files, rather than \texttt{.f} as in the public version. Also, a compiler bug causing segmentation faults forced us to compile a small portion of \bhac with the option \texttt{-hipa0}, switching off procedural optimization for a few kernels. This issue has been notified, and will be fixed in future versions of the compiler (from v.12.0 on). All the other kernels have simply been compiled with \texttt{-O3} optimization.

\begin{figure}[!]
\begin{center}
\includegraphics[width=\columnwidth]{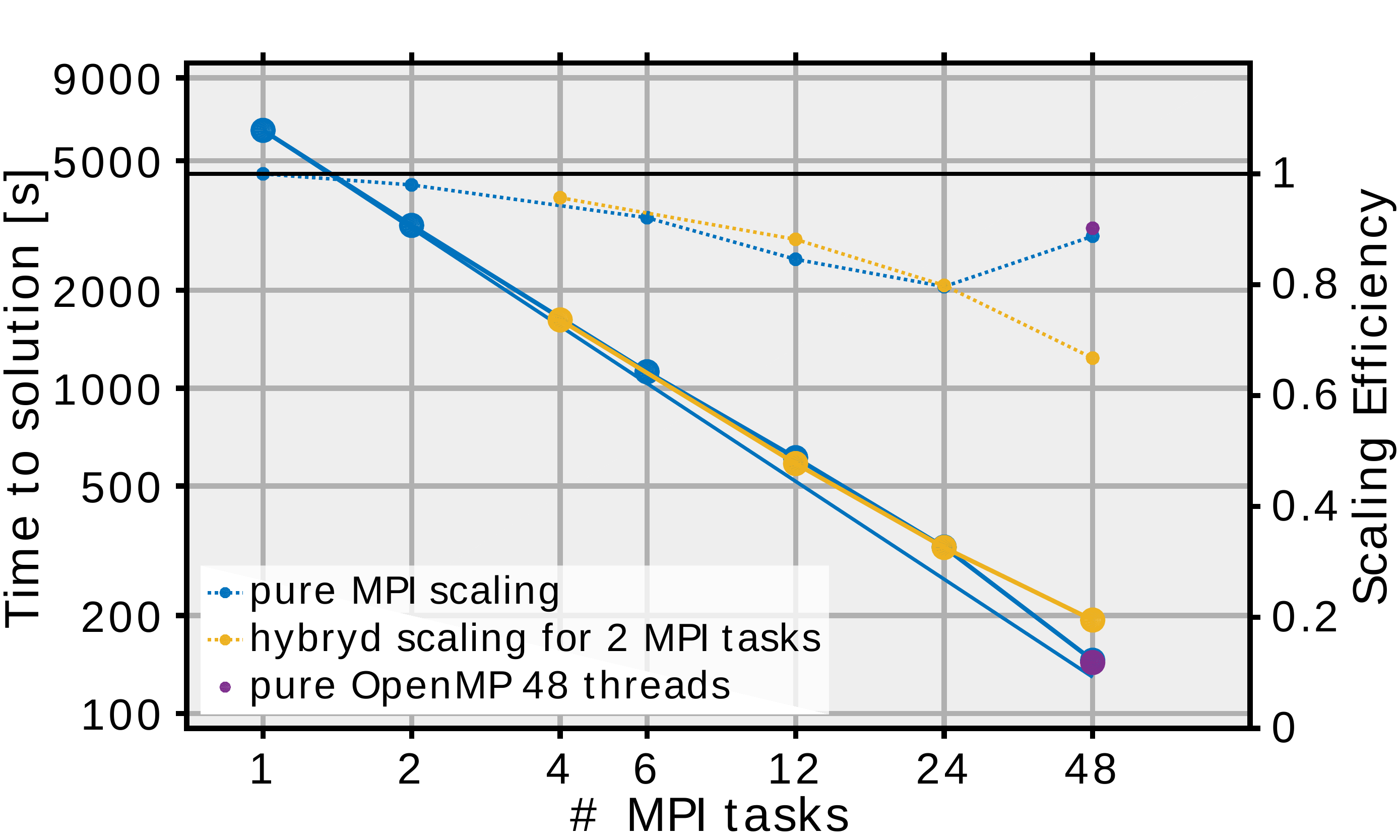}
\caption{Out-of-the-box scaling of the single node test on the A64FX node part of LRZ's \emph{BEAST} cluster. The nodes shows excellent scaling, especially for both pure paradigms, although overall inferior performance (about 3x slower than a single SuperMUC-NG node). Dedicated tuning of the application is likely to bring major improvement, but is outside our current scope.}
\label{fig:beast}
\end{center}
\end{figure}

We then ran the code interactively on a single node, performing a strong-scaling test similar to what is shown in Figure \ref{fig:single-node} (for the same code version). We tested pure MPI scaling from 1 to 48 cores (the maximum of the node; in blue), then a full-node pure OpenMP run (purple dot), and a hybrid configuration with 2 MPI tasks and increasing OpenMP threads (in yellow). Whenever running with OpenMP, we followed the advice given by the runtime to set \texttt{MV2\_ENABLE\_AFFINITY=0} to avoid potential thread overlap on individual physical cores; the results are displayed in Figure \ref{fig:beast} (median of 15 measures, statistical errors not shown as always less than 1\%). 

\change[sc]{While the scaling efficiency is satisfactory,} 
{A comparison with Figure} \ref{fig:single-node} shows that
the absolute performance is still about a factor 3 lower than on SuperMUC-NG. The pure paradigms perform best, with parallel efficiency around 90\% (and note how in the pure MPI scaling the efficiency is higher in the full node than for 12 or 24 cores). 
\add[sc]
{These efficiency values appear well inline with those of SuperMUC-NG, though slightly higher for the pure paradigms, slightly lower for the hybrid run with 2 MPI tasks}.

Inspection of the assembly code with the \texttt{perf record} and \texttt{perf report} tools showed very poor usage of the SVE set. The SIMD version of some instructions are used, although in limited amount when compared to SuperMUC-NG. Considering the high degree of SIMD vectorization of BHAC, even in its baseline version, the missing performance is not surprising. Additional reasons may include the fact that the node and compiler were never before tested on FORTRAN applications (otherwise administrators and expert users may 
\change[sc]{provide}
{have provided} guidelines for compiler optimization), and the aforementioned procedural optimization switch off: even though limited to a few kernels, those concern a rather important part of the grid tree management. 

Finally, the measured performance of \bhac seems in line with other applications (e.g. benchmarks or other code optimization projects). The emerging pattern is that high speedups typically require much more involved optimization work, such as explicit loop unrolling and development of detailed performance models \cite{alappat2020performancemodel}, which call for dedicate projects, when not dedicated staff; but can result in general optimization hints all users will benefit from. 

We thus consider our test run concluded, having answered our most pressing question (i.e., astrophysical real-world applications optimized for x86 SIMD are unlikely to benefit from SVE out-of-the-box), and having laid the basis for porting BHAC to A64FX; future studies may start from here. The result is undesirable, but hopefully software/hardware co-design efforts will alleviate this discrepancy in the near future.

\section{Conclusions and Outlook}\label{s:end}

In presenting our work on the \bhac code, we showcase some HPC tools and techniques of key importance in the optimization of codes' threading efficiency, especially in the case of newly introduced OpenMP parallelization layers aimed at production on modern supercomputers.

A characterization of the \bhac code infrastructure has been obtained using single node, 16 nodes and 800 node scenarios 
\add[sc]{on the Intel Xeon based SuperMUC-NG supercomputer}. The setup used for this initial investigation was deliberately chosen as simple as possible: we have resorted to non-AMR grids and IO has also been excluded. In this simple scenario, OpenMP correctness was addressed and imbalance was solved through \texttt{dynamic} scheduling (remaining OpenMP potential gain: $0.5\%$ as reported by VTune).

Guided by VTune's hotspots and threading analyses and APS' communication pattern report, the OpenMP serial fraction was significantly reduced, leading to a performance increase of $27\%$ at scale. 
\add[sc]
{We observed that the MPI node memory management was one of the main reason hindering the scale-up of the OpenMP optimizations to larger node numbers; we got the best results by switching Intel MPI's shared memory intra-node communication off.}
While there remains room for improvement, the project has shown that the hybrid implementation is capable to efficiently utilize over $30\, 000$ cores allowing to study large scale problems in scientific production.
The optimizations performed here have been added to the v1.1 release of \bhac \footurl{https://gitlab.itp.uni-frankfurt.de/BHAC-release/bhac/-/releases/v1.1}.  
\add[sc]
{We finally ported the code to a different architecture, an ARM A64FX vector processor part of the LRZ BEAST cluster. We observed with disappointment that the threading and SIMD optimizations (achieved in the course of several projects of which this work is only the last one) have little impact on the usage of vector instructions and thus on the final node-level performance, which remains about factor of 3 lower than on the x86 machine.

}

The improvements 
\change[sc]{made through in the AstroLab project}
{described in this work} are already merged into the staging branch of \bhac and will become part of the next public release.

\change[sc]{There are several potential next steps:}
{The modernization work on \bhac is however not concluded. Among the main issues that remain open for further projects we can list:}
\textit{1.} addressing the remaining OpenMP serial fraction in case of mixed OpenMP/MPI parallelization.  
\textit{2.} Extending the optimizations to the AMR case.
\textit{3.} Investigating the performance for a production level setup using AMR.  Here, next to the issues due to OpenMP/MPI hybridization, potential load-imbalance on the MPI level will need to be addressed.  

\section*{Acknowledgments}
S.C. thanks Dr. Josef Weidendorfer, leader of LRZ Future Computing Group, for the support on the test runs with the A64FX nodes described in Section \ref{s:beast}.

\bibliographystyle{hide/elsarticle-num-names}
\bibliography{astro.bib}

\end{document}